\documentclass[tighten,twocolumn]{aastex63}

\usepackage{graphicx}
\usepackage{ulem}
\usepackage{amsmath}
\usepackage{wrapfig}
\usepackage[thinlines]{easytable}
\usepackage{tikz}
\usepackage{hyperref}

\newcolumntype{P}[1]{>{\centering\arraybackslash}p{#1}}

\begin{document}

\shorttitle{}
\shortauthors{}

\correspondingauthor{Fulya K{\i}ro\u{g}lu}
\email{fulyakiroglu2024@u.northwestern.edu}

\author[0000-0003-4412-2176]{Fulya K{\i}ro\u{g}lu}
\affiliation{Center for Interdisciplinary Exploration \& Research in Astrophysics (CIERA) and Department of Physics \& Astronomy \\ Northwestern University, Evanston, IL 60208, USA}

\author[0000-0002-4086-3180]{Kyle Kremer}
\affiliation{Department of Astronomy \& Astrophysics, University of California, San Diego; La Jolla, CA 92093, USA}

\author[0000-0002-7132-418X]{Frederic A. Rasio}
\affiliation{Center for Interdisciplinary Exploration \& Research in Astrophysics (CIERA) and Department of Physics \& Astronomy \\ Northwestern University, Evanston, IL 60208, USA}

\title{Beyond Hierarchical Mergers: Accretion-Driven Origins of Massive, Highly Spinning Black Holes in Dense Star Clusters}

\begin{abstract}

GW231123, the most massive binary black hole (BBH) merger detected by LIGO/Virgo/KAGRA, highlights the need to understand the origins of massive, high-spin stellar black holes (BHs). Dense star clusters provide natural environments for forming such systems, beyond the limits of standard massive star evolution to core collapse. While repeated BBH mergers can grow BHs through dynamical interactions (the so-called ``hierarchical merger'' channel), most star clusters with masses $\lesssim 10^6\,M_\odot$ have escape speeds too low to retain higher-generation BHs, limiting growth into or beyond the mass gap. In contrast, BH--star collisions with subsequent accretion of the collision debris can grow and retain BHs irrespective of the cluster escape speed.
Using $N$-body (\texttt{Cluster Monte Carlo}) simulations, we study BH growth and spin evolution through this process and we find that accretion can drive BH masses up to at least $\sim200\,M_\odot$, with spins set by the details of the growth history. BHs up to about $150\,M_\odot$ can reach dimensionless spins $\chi \gtrsim 0.7$ via single coherent episodes, while more massive BHs form through multiple stochastic accretion events and eventually spin down to $\chi \lesssim 0.4$. These BHs later form binaries through dynamical encounters, producing BBH mergers that contribute up to $\sim10\%$ of all detectable events, comparable to predictions for the hierarchical channel. 
However, the two pathways predict distinct signatures: hierarchical mergers yield more unequal mass ratios, whereas accretion-grown BHs preferentially form near-equal-mass binaries. The accretion-driven channel allows dense clusters with low escape speeds, such as globular clusters, to produce highly spinning BBHs with both components in or above the mass gap, providing a natural formation pathway to GW231123-like systems.
\vspace{1cm}
\end{abstract}

\section{Introduction}

A number of gravitational wave (GW) detections by the LIGO-Virgo-KAGRA (LVK) collaboration have provided direct evidence for the existence of BHs in the so-called “upper mass gap” from roughly $40$-$120\,M_{\odot}$, expected from pair instabilities \citep[e.g.,][]{Woosley_2017}.
Dense star clusters offer multiple pathways for the formation of these massive BHs. One such route involves the formation of very massive stars through repeated stellar collisions in clusters with high central densities within the first few Myr, which then collapse to form black holes (BHs) within or beyond the upper mass gap \citep[e.g.,][]{mapelli2016,DiCarlo_2019,Kremer_2020,Gonzalez_2021}. This scenario requires the collision rate to be high enough that the time between collisions is shorter than the typical stellar lifetime. In extreme cases, runaway collisions can lead to the formation of even more massive BHs, potentially linking to the intermediate-mass BH regime, with masses in the range of $\sim 10^2$–$10^4\,M_\odot$ \citep[e.g.,][]{Gurkan_2004,PortegiesZwart2004,Haberle_2024,Gonzalez_2024}.

After forming via core-collapse, stellar-mass BHs can also grow through repeated mergers with other BHs as they segregate toward the cluster center and dynamically pair up \citep[e.g.,][]{MillerHamilton2002,GerosaBerti2017,Rodriguez_2019,Antonini_2019,FragioneRasio2023}. The maximum BH growth through this channel depends critically on the recoil of merging BBHs through
asymmetric GW emission. The magnitude of the recoil is highly sensitive to the spin magnitudes and mass ratio of the merging components
\citep[e.g.,][]{Merritt_2004,Campanelli_2007,Berti_2008,Lousto_2012,Gerosa_2019}. In general, higher spins produce larger recoil velocities, increasing the likelihood that the remnant is ejected from the cluster; even moderate first-generation spins (e.g., $\chi \approx 0.2$) can substantially suppress the rate of second-generation BBH mergers, particularly those involving BHs in the pair-instability mass gap \citep{Rodriguez_2019}. 
Hence, in less massive clusters with lower escape speeds, the recoil from merging BBHs can quickly quench the growth process \citep{OLeary_2006,Holley-Bockelmann_2008,Fragione_2018}. In contrast, massive star clusters---with escape speeds in excess of roughly $ 1000\,\rm{km\,s^{-1}}$---are more likely to retain merger remnants, making continued BH growth much more probable \citep{Antonini_2016,Antonini_2019,Rodriguez_2020,Fragione_2022,Mai2025}. However, massive BHs produced through repeated mergers are expected to have low spins as a consequence of the random orientations of successive merger events \citep[e.g.,][]{Hughes_2003}.

BHs can also grow via accretion of gas, which can occur in a variety of astrophysical contexts \citep[e.g.,][]{Ostriker1983,vesperini2010,McKernan2012,Leigh2013,Lupi2016,Bartos_2017,Shi2024,Roupas2025}. In dense star clusters, physical collisions and tidal disruptions offer natural pathways for accretion \citep[e.g.,][]{Giersz_2015,Hellstrom2022,Rose_2022,Kiroglu_2025a}. In old globular clusters, this mechanism can be reasonably neglected, as most BH–star collisions occur after the most massive stars (including stellar-merger products) have evolved off the main sequence and are therefore unlikely to add substantial mass to the BH \citep[e.g.,][]{Kremer2019_tde,Kremer_2022}. However, in a pair of recent studies \citep{Kiroglu_2025a,Kiroglu_2025b}, we show that the high central densities of young massive star clusters, combined with elevated binary fractions among massive stars \citep[e.g.,][]{Sana2025}, significantly enhance the collision rate of BHs with massive stars. In particular, we showed that these collisions can affect over $30\%$ of BHs within the first $100\,$Myr of cluster evolution, suggesting they may play a substantial role in shaping the BH population.

Modeling BH--star collisions is challenged by two major uncertainties: the amount of stellar mass captured by the BH during the disruption itself and the subsequent accretion phase. Following the disruption of the star, a fraction of the stellar debris becomes bound to the BH with the captured fraction depending on the mass ratio and the strength of the encounter. In some instances, the full disruption process may feature multiple passages \citep[e.g.,][]{Kremer_2022,Kiroglu_2023,Vynatheya_2024}. Subsequent accretion of bound material depends on multiple factors including the geometry of the accretion flow (e.g., a disk versus a quasi-spherical envelope), potential mass loss from disk winds, and accretion feedback \citep[for further detail, see][]{Kremer_2022,Kremer_2023}. Collectively, these many factors are critical in shaping the mass and spin distributions of BHs undergoing these stellar collisions, and  particularly those BHs that go on to form GW sources.

In this work, we investigate in detail the formation of massive BHs through close encounters between BHs and stars, including mergers and collisions, in dense star clusters. We implement, for the first time, BH mass growth and spin evolution from stellar collisions directly within our $N$-body simulations. While fully capturing the outcome of BH--star close encounters requires detailed hydrodynamic calculations, we account for this uncertainty by varying the accretion efficiency between $0\%$ and $100\%$ in our models, bracketing the range of possible outcomes.

Our paper is organized as follows. In Section \ref{sec:models}, we summarize the methods used to model BH formation and evolution in dense star clusters. Section~\ref{sec:results} presents the results of our $N$-body simulations, focusing on the rates and efficiencies of the processes driving BH mass growth. We further investigate how stellar accretion shapes the spin evolution of BHs and, consequently, the demographics of merging BBHs, comparing our predictions with the most recent GWTC-4 events. Finally, we conclude and discuss our findings in Section~\ref{sec:conclusion}.

\section{Method}

\label{sec:models}

We perform 12 cluster simulations using the \texttt{Cluster Monte Carlo} (\texttt{\texttt{CMC}}) code, a H\'{e}non-style $N$-body code for stellar dynamics \citep[see][for a detailed review]{Rodriguez_2022}. \texttt{\texttt{CMC}} incorporates various physical processes essential for studying both formation and evolution of stellar-mass BHs, including stellar and binary star evolution using the \texttt{COSMIC} population synthesis package \citep{Breivik2020} which includes our most up-to-date understanding of the formation of compact objects, including prescriptions for natal kicks, mass-dependent fallback, and (pulsational) pair-instability supernovae \citep{Fryer_2001,Belczynski_2002}, three-body binary formation \citep{Morscher_2015}, and direct integration of small-$N$ resonant encounters \citep{FregeauRasio2007} including post-Newtonian effects \citep{Rodriguez_2018}.

\begin{table}[t ]
\startlongtable
\begin{deluxetable*}{l|cc|ccc|cccc|c}
\tablewidth{0pt}
\tablecaption{List of Cluster Models
\label{table:mergers}}
\tablehead{
    \multicolumn{3}{c}{} &
    \multicolumn{3}{c}{\# BH interactions} &
	\multicolumn{4}{c}{\# BHs $ (M > 40.5\,M_{\odot})$}\\
	\colhead{$^{1}$Model} &
     \colhead{$^{2}r_v/\rm{pc}$} &
    \colhead{$^{3}f_{\rm acc}$} &
    \colhead{$^{4}$BH--star coll} &
    \colhead{$^{5}$BH--star merger} &
    \colhead{$^{6}$BH-BH} &
     \colhead{$^{7}$star-star} &
    \colhead{$^{8}$BH--star} &
	\colhead{$^{9}$BH-BH} &
 	\colhead{$^{10}$Total} & 	
    \colhead{$^{11}$Max BH mass}
	} 
\startdata
1a & 1 & 0  & 16 & 29 & 80 & 65 & 0 & 24 & 89 & 123 \\
1b & 1 & 0  & 11 & 36 & 64 & 66 & 0 & 34 & 100 & 125\\ 
2a & 0.5 & 0  & 66 & 6 & 123 & 26 & 0 & 41 & 67 & 133 \\
2b & 0.5 & 0 &  60 & 7 & 137 & 17 & 0 & 53 & 70 & 166 \\
\hline
3a & 1 & 0.5  & 13 & 39 & 91 & 54 & 0 & 36 & 90 & 125 \\
3b & 1 & 0.5  & 20 & 37 & 76 & 61 & 0 & 29 & 90 & 152 \\
4a & 0.5 & 0.5 & 57 & 4 & 134 & 21 & 14 & 46 & 81 & 180 \\
4b & 0.5 & 0.5 & 50 & 10 & 114 & 20 & 6 & 44 & 70 & 193 \\
\hline
5a & 1 & 1 & 17 & 44 & 80 & 57 & 1 & 25 & 83 & 119 \\
5b & 1 & 1  & 24 & 35 & 89 & 56 & 2 & 32 & 90 & 123 \\ 
6a & 0.5 & 1  & 127 & 6 & 127 & 25 & 30 & 41 & 96 & 330 \\
6b & 0.5 & 1  & 122 & 2 & 118 & 29 & 24 & 47 & 100 & 467 \\
\hline
\enddata
\vspace{0.5cm}
\tablecomments{Complete list of all cluster simulations performed in this study. For each set of cluster initial conditions, we perform two independent realizations, denoted ‘a’ and ‘b’ for each model number. Columns 2 and 3 list the initial virial radius ($r_v$) and accretion efficiency ($f_{\rm acc}$) in each simulation. Columns 4–6 report the number of BH–star collisions, BH–star mergers, and BH–BH mergers, respectively. Column 7 lists the number of mass-gap BHs ($M_{\rm BH} > 40.5\,M_{\odot}$) formed via stellar collisions prior to core-collapse supernovae. Columns 8 and 9 show the number of mass-gap BHs formed through accretion during all BH–star interactions and through BH–BH mergers, respectively. The final column (11) gives the mass of the most massive BH formed in each simulation. 
}
\label{table:models}
\end{deluxetable*}
\end{table}

All models we consider here assume an initial population of $N=8\times10^5$ objects, including single and binary stars. Initial stellar masses  are drawn from an initial mass function (IMF) ranging from $0.08–150\,M_{\odot}$, following slopes of \cite{Kroupa2001}. Each model is initially described by King profiles \citep{King1962} with a fixed concentration parameter of $W_0=5$. We adopt a fixed metallicity $Z=0.1\, Z_{\odot}$ and galactocentric distance of $20 \,$kpc in a Milky Way–like galactic potential. We also vary the initial cluster virial radius: $r_v=[0.5,1]\,$pc.
All these models have escape velocities similar to typical globular clusters, ranging from 10 to $ 100\,\rm{km\,s^{-1}}$.

We assume an initial low-mass ($<15\, M_{\odot}$) binary fraction of $5\%$ in all models. The initial binary fraction for massive stars ($\geq 15\,M_{\odot}$) stars is set to $100\%$. For low-mass binaries, primary masses are drawn randomly from our IMF. Secondary masses are drawn assuming a flat mass ratio distribution in the range [0.1, 1], and initial orbital periods are drawn from a log-uniform distribution $dn/ d \log P \propto P$.
For the secondaries of the massive stars $(>15 \,M_{\odot})$, a flat mass ratio distribution in the range [0.6, 1] is assumed, and initial orbital periods are drawn from the distribution $dn /d \log P \propto P^{-0.55}$ \citep[e.g.,][]{Sana_2012}. For all binaries, binary semi-major axes are drawn from near contact to the hard/soft boundary, which is determined using the local velocity dispersion, and initial eccentricities are drawn from a thermal distribution. 

We employ the standard $\alpha–\lambda$ formalism for common-envelope (CE) evolution following \citet{Hurley2002} within \texttt{COSMIC}. In this prescription, binaries undergoing unstable mass transfer enter a CE phase, which may be ejected through the injection of orbital energy. Here, $\lambda$ quantifies the binding energy of the envelope to the stellar core, while $\alpha$ represents the efficiency with which orbital energy is transferred to unbind the envelope. Consistent with the default implementation in \texttt{COSMIC}, we adopt a variable $\lambda$ that depends on the evolutionary state of the donor star, as described in \citet{Claeys2014}, and a constant CE efficiency parameter $\alpha = 1.0$ \citep[e.g.,][]{Nelemans2001,Dominik2012}. 
\vspace{0.9cm}
\subsection{Black Hole Formation}

Stellar-mass BHs form as the evolutionary endpoints of massive stars, with their final masses set largely by two uncertain factors: the progenitor’s pre-collapse mass (and core–envelope structure) and the physics of the subsequent supernova. The pre-collapse mass, in turn, is strongly regulated by metallicity-dependent stellar winds \citep[e.g.,][]{Vink_2001}. We adopt the \citet{Fryer_2012} delayed supernova-engine prescription for BH formation in our stellar evolution modeling. This choice, which is the current default in \texttt{CMC}, yields a robust population of BHs in the lower-mass gap ($\approx 2–5\,M_{\odot}$), consistent with observations from the third
GW Transient Catalog \citep{Abbott_2023}.

In our models, we incorporate prescriptions for pulsational pair-instabilities and pair-instability supernovae, following \citet{Belczynski_2016}. Specifically, we assume that stars with pre-collapse helium core masses between 45 and 65 $M_{\odot}$ undergo pulsations that expel large amounts of mass, ultimately reducing the core mass to no more than  $45\,M_{\odot}$. Accounting for a $10\%$ loss in mass from neutrino mass loss during collapse, the resulting BH has a mass of approximately  $40.5\,M_{\odot}$. 

For stars with helium core masses exceeding $65\,M_{\odot}$, we assume complete disruption in a pair-instability supernova, leaving no remnant. However, dynamical interactions enable the formation of BHs with masses exceeding $40.5\,M_{\odot}$. Repeated mergers of massive stars can produce unusual core-envelope mass ratios, allowing them to bypass the typical pulsational pair-instability limits. Clusters with primordial mass segregation, fractal initial conditions, or elevated binary fractions indeed have been shown to favor the growth of massive BHs through stellar collisions \citep[e.g.,][]{DiCarlo_2019,Kremer_2020,Gonzalez_2021}.

We assume that all BHs born via stellar collapse are born non-spinning \citep{Fuller_2019} and receive no natal kicks. BHs can subsequently acquire spin through mergers with other BHs or by accreting mass from non-degenerate stars. Below, we detail the prescriptions used to model the evolution of BH mass and spin following both stellar accretion and compact object mergers.

\subsection{Black Hole Mergers}

\begin{figure*}
    \centering
    \includegraphics[width=0.8\linewidth]{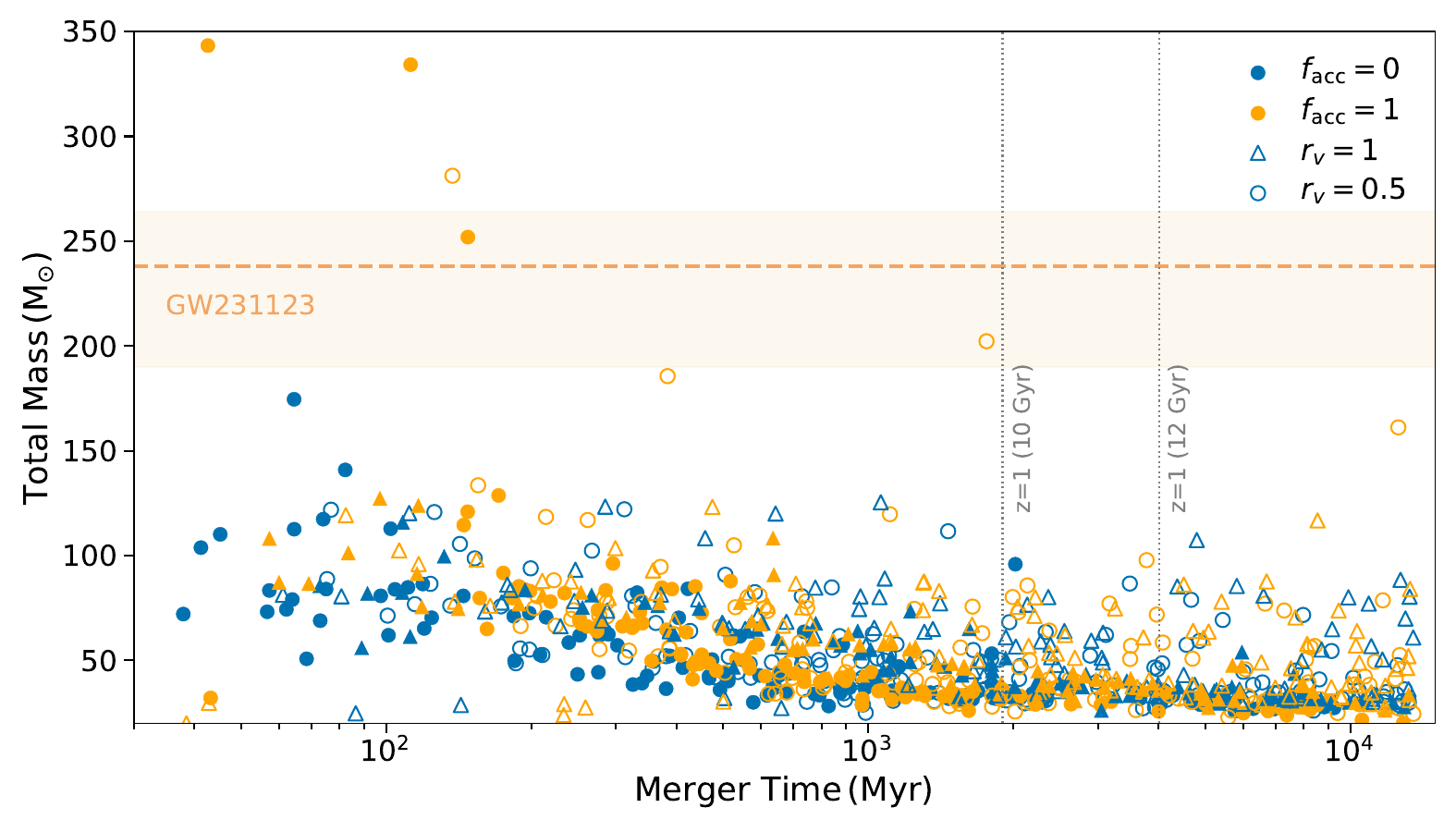}
    \caption{\footnotesize Total mass of BBH mergers as a function of merger time without accretion ($f_{\rm acc} = 0$, blue) and with accretion ($f_{\rm acc} = 1$, orange) in globular clusters with an escape speed of about $ 50\,\rm{km\,s^{-1}}$.
 Solid and open circles indicate in-cluster and ejected mergers, respectively. Triangles and circles denote models with initial virial radii of $r_{\rm v} = 1.0$~pc and $r_{\rm v} = 0.5$~pc, respectively. The shaded orange band marks the mass range of GW231123 \citep{GW231123}, with the dashed line indicating its median total mass. The dotted gray lines indicate the redshift $z=1$, assuming present-day cluster ages of 10 Gyr and 12 Gyr.}
        \label{fig:mbbh_total}
\end{figure*} 
Once formed (either dynamically or as a primordial pairing at star formation), BH binaries can inspiral and eventually merge due to GW emission, as they undergo repeated hardening interactions in the dense core of the cluster.
In the event of a BBH merger, we compute the spin, mass and GW recoil kick of the new BH using the method described in \cite{Rodriguez_2018}, which in turn implements phenomenological fits to numerical and analytic relativity calculations \citep{Campanelli_2007,Gonzalez_2007,Barausse_2009,Lousto_2013}. While the merger of two non-spinning, equal-mass BHs produces a remnant with $\chi \simeq 0.69$, the typical spin magnitudes of 2G BHs in our models are in the range $\chi \approx 0.6$–$0.8$.

\newcommand{\vta}[1]{\vert \boldsymbol{\chi}_{#1}        \vert}
\newcommand{\vtaf}  {\vert \boldsymbol{a}_{\rm fin}   \vert}

\newcommand{\chip}  {\vert \boldsymbol{\chi}_{ 1}   \vert}
\newcommand{\chis}  {\vert \boldsymbol{\chi}_{ 2}   \vert}

\newcommand{\Sf}  {\vert \boldsymbol{S}_{ \rm f}   \vert}

\newcommand{\chif}  {\vert \boldsymbol{\chi}_{ \rm f}   \vert}

\newcommand{\vtl}   {\vert \boldsymbol{       {\ell}} \vert}

\subsection{Black Hole Accretion}

We allow BHs to grow via accretion during close encounters with stars, applying the same $f_{\rm acc}$ efficiency factor to both BH--star collisions modeled in \texttt{Fewbody} and BH–star mergers in \texttt{COSMIC}, to ensure consistency in our treatment of BH mass growth. Following \citet{Fabian1975}, we define the tidal disruption radius as
\begin{equation}
\label{eq:rt}
r_T = f_{\rm p}\,\left(\frac{M_{\rm BH}}{M_{\star}}\right)^{1/3}\,R_{\star},
\end{equation}
where $M_{\rm BH}$ is the BH mass, $M_{\star}$ and $R_{\star}$ are the stellar mass and radius, respectively, and $f_p$ is a dimensionless factor that depends on the star’s internal structure.

Accurately resolving the outcomes of close BH–star encounters requires detailed hydrodynamic simulations \citep[e.g.,][]{Kremer_2022, Ryu_2022, Kiroglu_2023}, which are computationally infeasible within the \texttt{\texttt{CMC}} framework. To make the problem tractable, we adopt a simplified prescription: stars are assumed to follow parabolic orbits and are fully disrupted on their first pericenter passage if $r_p \leq r_T$, where $r_T$ is evaluated using Equation~(\ref{eq:rt}) with $f_p = 1$. We further assume that the entire stellar mass remains gravitationally bound to the BH following disruption \citep{Kremer_2022}, and that the total mass ultimately accreted by the BH is given by $M_{\star} \times f_{\rm acc}$. In this work, we explore models with accretion efficiency values $f_{\rm acc} = [0, 0.5, 1]$. 
We assume that accretion by the BH occurs immediately after the interaction, i.e., that the accretion timescale is short compared to any other relevant timescale. The actual accretion timescales are highly uncertain and are determined by the viscous timescale of the resulting accretion disk, which can range from days to years depending on the geometry and physical sources of viscosity \citep{Kremer_2022}.

We calculate the change of the BH spin magnitude through the accretion of disrupted material as described in \cite{Bardeen_1972}, where we assume that the final BH will have a mass and angular momentum nearly equal to those of the binary system at the innermost stable circular orbit, $r_{\rm ISCO}$. Assuming the accretion disk lies in the equatorial plane of the BH and that material is accreted directly from the innermost stable circular orbit (ISCO), the final dimensionless spin parameter of a BH with final mass $M_{\rm f}$ is given, for $M_{\rm f}/M \leq r_{\rm ISCO}^{1/2}$, by \cite{Bardeen_1972,Volonteri_2013}
\begin{equation}
\label{eq:chi}
    \chi_{\rm f} =
    \frac{r_{\rm ISCO}^{1/2}}{3}\frac{M}{M_{\rm f}}\left[4-\left(3 \left(\frac{M}{M_{\rm f}}\right)^2 r_{\rm ISCO}-2\right)^{1/2}\right] 
\end{equation}
while  $\chi_{\rm f} =1$ for $M_{\rm f}/M \geq \it{r_{\rm ISCO}}^{\rm{1/2}}$, where
\begin{align}
r_{\rm{ISCO}} &= 3 + Z_2 - \rm{sign}(\chi) \sqrt{(3-Z_1)(3+Z_1+2Z_2)} \label{eq:R_isco}\\
Z_1 &\equiv 1 + (1-\chi^2)^{1/3}\left[(1+\chi)^{1/3} + (1-\chi)^{1/3}\right]\\
Z_2 &\equiv \sqrt{3 \chi^2 + Z_1^2}.
\end{align}

\begin{deluxetable*}{l|cccc|cccc|cccc}
\tablewidth{0pt}
\tablecaption{List of Binary Black Hole Mergers
\label{table:mergers}}
\tablehead{
    	\colhead{$^{1}$Model} &
    \multicolumn{4}{c}{\# All BBH Mergers} &
	\multicolumn{4}{c}{\# Mass-Gap Primary}&
    \multicolumn{4}{c}{\# Mass-Gap Secondary}\\
	\colhead{} &
    \colhead{$^{2}$All} &
    \colhead{$^{3}$1G} &
    \colhead{$^{4}$2G} &
     \colhead{$^{5}$Accreted} &
    \colhead{$^{6}$All} &
    \colhead{$^{7}$1G} &
    \colhead{$^{8}$2G} &
     \colhead{$^{9}$Accreted} &
        \colhead{$^{10}$All} &
    \colhead{$^{11}$1G} &
    \colhead{$^{12}$2G} &
     \colhead{$^{13}$Accreted}
	} 
\startdata
1a & 80 & 75 & 5 & 0 & 23 & 18 & 5 & 0 & 8 & 8 & 0 & 0 \\
1b & 64 & 57 & 7 & 0 & 24 & 17 & 7 & 0 & 6 & 5 & 1 & 0 \\
2a & 123 & 104 & 19 & 0 & 23 & 10 & 13 & 0 & 5 & 4 & 1 & 0 \\
2b & 137 & 120 & 17 & 0 & 20 & 7 & 13 & 0 & 5 & 3 & 1 & 1 \\
\hline
3a & 91 & 70 & 14 & 7 & 26 & 13 & 12 & 1 & 11 & 11 & 0 & 0 \\
3b & 76 & 69 & 4 & 3 & 20 & 15 & 4 & 1 & 5 & 4 & 1 & 0 \\
4a & 134 & 104 & 18 & 12 & 22 & 4 & 13 & 5 & 8 & 6 & 1 & 1 \\
4b & 114 & 94 & 11 & 9 & 23 & 9 & 8 & 6 & 1 & 1 & 0 & 0 \\
\hline
5a & 80 & 62 & 7 & 11 & 27 & 12 & 5 & 10 & 6 & 5 & 0 & 1 \\
5b & 89 & 68 & 5 & 16 & 22 & 9 & 3 & 10 & 11 & 7 & 0 & 4 \\
6a & 127 & 90 & 14 & 23 & 28 & 6 & 9 & 13 & 7 & 2 & 0 & 5 \\
6b & 118 & 86 & 11 & 21 & 24 & 5 & 8 & 11 & 7 & 2 & 0 & 5 \\
\hline
\enddata
\vspace{0.3cm}
\tablecomments{Summary of BBH mergers across all cluster models. Columns 2–5 list the total number of BBH mergers, including those with 1G, 2G and accreted primary BHs. Columns 6–9 (10–13) show BBH mergers with primaries (secondaries) within or above the mass gap, subdivided similarly. Mass-gap 1G BHs originate from the collapse of very massive stars formed through stellar collisions.}
\end{deluxetable*}

Spin evolution depends on the angular momentum of the accreted material: coherent (aligned) accretion spins up the BH, while counter-rotating accretion spins it down. In \texttt{CMC}, we assume equal probabilities for prograde and retrograde accretion. However, retrograde accretion is more efficient at reducing spin due to the larger ISCO radius \citep[e.g.,][]{Hughes_2003}. Because our spin evolution prescription is independent of the spin–tilt angle, it effectively provides upper and lower limits on the resulting BH spin. In reality, stellar orbits in clusters are randomly oriented, and disk–spin misalignments are expected to occur with a broad distribution. This could change the details of the resulting $\chi$ distribution but is unlikely to affect our results qualitatively, as we do include both limiting cases of aligned and anti-aligned disks.
\vspace{0.4cm}
\section{Results}
\label{sec:results}

\subsection{Black Hole Growth}

Table~\ref{table:models} summarizes the types of BH interactions and the formation channels of massive BHs in our suite of 12 cluster simulations, which span a range of cluster densities (parameterized by the initial $r_v$; column 2 in the table) and accretion efficiencies (column 3). Each model forms and retains approximately $10^3$~BHs via collapse of massive stars. After formation, a subset of these BHs undergo further growth by merging with other BHs (BH–BH, column 6) or through interactions with stars—either via direct physical collisions during a hyperbolic encounter or by merger in a bound binary (i.e., a failed CE). We separate these BH–star interactions into two categories in columns 4 and 5, corresponding to BH--star collisions and BH--star mergers, respectively. Column 4 lists BH–star collisions that occur during few-body encounters, together with those arising from single–single interactions. Overall, more than $50\%$ of all BH–star collisions originate from single–single encounters, while at most $\approx 10\%$ occur during binary–binary interactions.

We find that a comparable fraction—about $10\%$ of the retained BHs—undergo BH–BH mergers (column 6), while another $10\%$ interact with stars (columns 4 and 5 combined). Comparing columns 4 and 5 for models with different $r_v$ we see that in compact clusters ($r_{\mathrm{v}} = 0.5\,$pc), BHs primarily gain mass through dynamical BH–star collisions, which occur frequently due to short encounter timescales; in more extended clusters ($r_{\mathrm{v}} = 1\,$pc), longer collision timescales allow more BH–star binaries to evolve and merge via stellar evolution. 
\begin{figure*}
\centering    \includegraphics[width=0.7\linewidth]{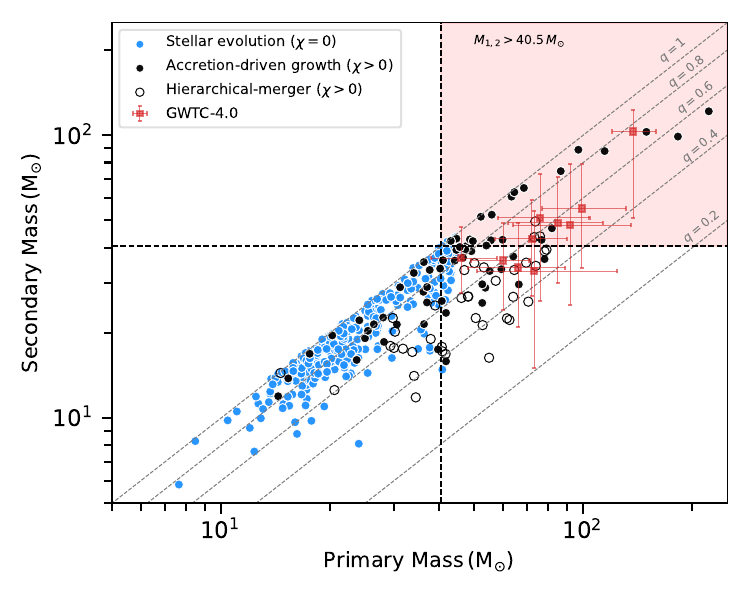}
    \caption{\footnotesize  Secondary versus primary component masses for all BBH mergers in our simulations with $f_{\rm acc}=1$. Blue circles show binaries with both components non-spinning, while black circles indicate binaries with at least one spinning BH. Filled black circles indicate BHs that have grown and spun up through accretion ($\chi > 0$), while open black circles represent second-generation (2G) BHs with a typical spin of $\chi \approx 0.7$. Red points with error bars show the GWTC-4 events with at least one component within or above the pair-instability mass gap \citep{LVK_2025}, with several lying in the top-right region where both BHs are in or above the gap $M_{1,2}>40.5\,M_{\odot}$. Gray dashed lines indicate constant mass ratios $q=0.2$–$1$. Accretion-driven growth populates the red shaded region, where both BHs lie within or above the mass gap and have nearly equal mass ratios whereas hierarchical mergers preferentially produce systems with lower mass ratios ($q \approx 0.5$).
}
        \label{fig:mass_comp}
\end{figure*}

Columns 7, 8, and 9 detail the different growth channels that produce BHs more massive than $40.5\,M_\odot$, which we adopt as the lower boundary of the pair-instability mass gap.
In models with an initial virial radius of $r_{\rm v} = 1\,\mathrm{pc}$, most mass-gap BHs originate from early collisions or mergers between massive stars \citep[the channel described in][]{DiCarlo_2019, Kremer_2020,Gonzalez_2021}. These stellar interactions typically occur within the first $\sim 5\,\mathrm{Myr}$ and often involve stars initially placed in binaries (with $M > 15\,M_\odot$), which enhances their dynamical cross sections and collision likelihoods.
By contrast, in more compact clusters with $r_{\rm v} = 0.5\,\mathrm{pc}$, BH growth is dominated by BH–star and BH–BH interactions. Although the rate of star–star mergers is higher in these denser clusters, the number of BHs exceeding $40.5\,M_{\odot}$ formed through this channel is lower because many of the massive stellar merger products collide with BHs before they can undergo core collapse. This trend is reflected in the higher number of BHs that grow through BH–star interactions, as reported in Column 8 of Table~\ref{table:models}.

We find that all three pathways—star–star collisions, BH–star collisions, and BH–BH mergers—contribute nearly equally to the formation of mass-gap BHs if we allow BHs to accrete significantly ($f_{\rm acc}=1$).
Overall, these results demonstrate that, depending on the accretion efficiency, up to $\sim10\%$ of all BHs in young star clusters can grow into the pair-instability mass gap through a combination of dynamical interactions.
In Column 11, we list the most massive BH formed in each simulation. As $f_{\rm acc}$ increases from 0 to 1, BHs grow to significantly higher masses, with the most compact clusters producing massive BHs beyond $200\, M_\odot$. These results underscore that both the accretion efficiency during interactions and the cluster’s dynamical environment are key factors in determining the formation of massive BHs.

\subsection{Massive Binary Black Hole Mergers}
Massive BHs formed through stellar mergers and/or accretion efficiently acquire binary companions and subsequently merge via dynamical interactions, either within the cluster or after ejection. 
In Figure ~\ref{fig:mbbh_total}, we show the total mass of BBH mergers as a function of cluster age across all cluster models, comparing results for two different accretion efficiencies. In the $f_{\rm acc} = 1$ models, BHs undergo significant growth through accretion at early times, leading to BBH mergers with total masses exceeding $250\,M_\odot$—comparable to high-mass events such as GW231123.
As initially noted in \citet{Rodriguez_2018}, the mass distributions of in-cluster versus ejected mergers diverge markedly at late times. This distinction arises from the long delay times ($\sim\, $Gyr) associated with ejected BBHs, which are often formed early and reflect the BH mass distribution at birth and/or are shaped by early stellar collision/accretion events. In contrast, in-cluster mergers occur promptly after binary assembly. Consequently, ejected massive BBH mergers observed at late times may provide insight into the early growth history and stellar collision processes within young star clusters.

\begin{figure*}
    \centering
    \includegraphics[width=0.9\linewidth]{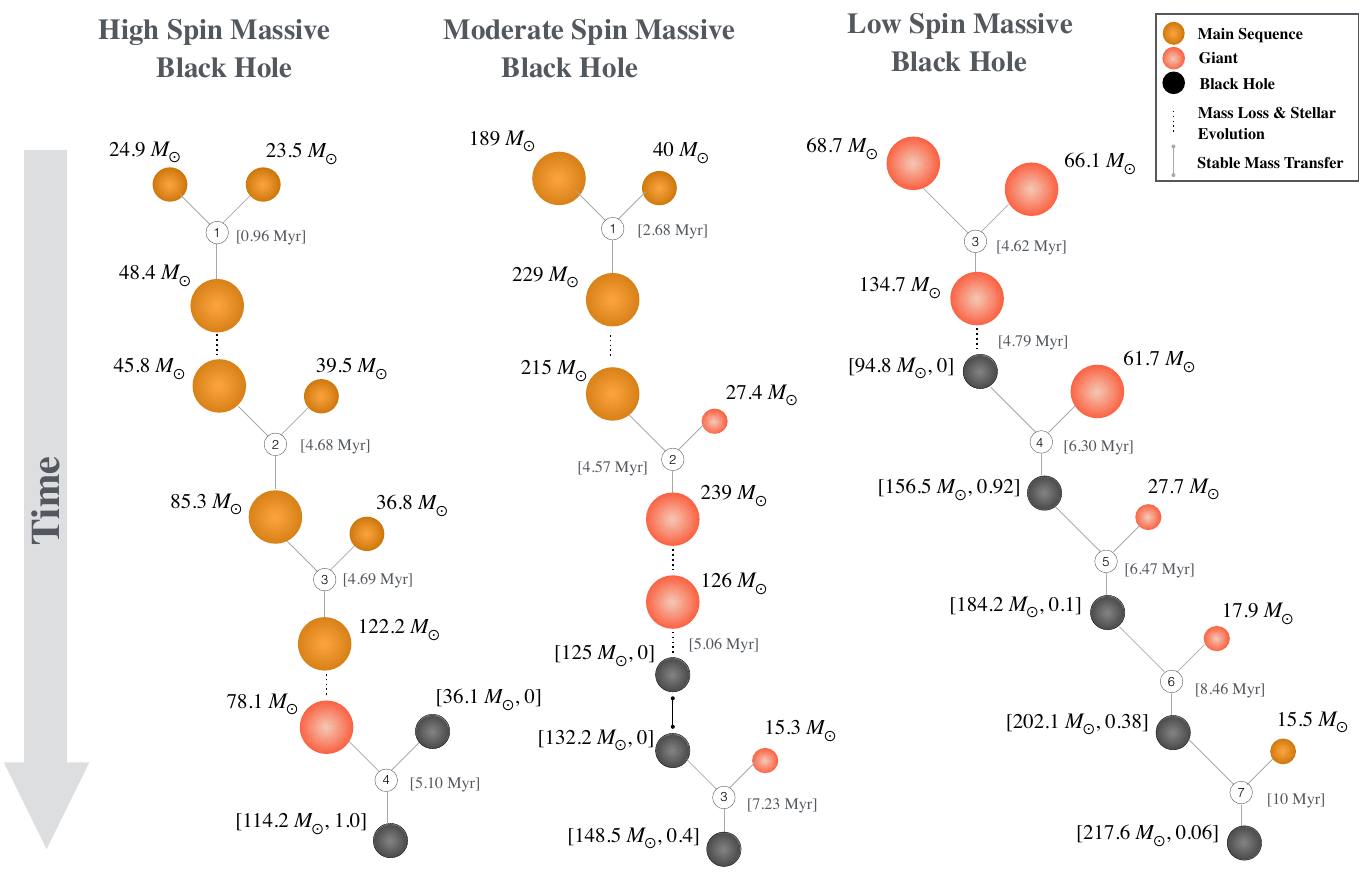}
    \caption{\footnotesize  Example collision histories illustrating the assembly of a massive BH through a sequence of mergers, shown for three representative cases of spin evolution. The mass of each object is labeled before and after each collision, with the collision time indicated in brackets. For BHs, spin values are given in parentheses alongside their masses. }
        \label{fig:coll_history}
\end{figure*}

Table~\ref{table:mergers} summarizes the number of BBH mergers in all cluster models, categorized by the evolutionary origin of their components. Column 2 lists all BBH mergers, including ejected and in-cluster, occurring within a Hubble time. Overall, the majority of BBH mergers involve first-generation (1G) BHs that are initially nonspinning, while up to $\sim20\%$ of mergers involve at least one spinning primary, driven by either stellar accretion or hierarchical BH mergers. Column 6 reports the number of BBH mergers in which the primary BH lies within or above the pair-instability mass gap. Nonspinning 1G BHs in the mass gap typically form from the direct collapse of massive stars created through stellar collisions. Across all models, the number of mass-gap mergers remains nearly constant at $\sim20$–$30$, comprising approximately $20\%$ of the total BBH merger population. This uniformity underscores the robustness of mass-gap BH mergers across a wide range of cluster densities and accretion efficiencies. By contrast, only a few percent of secondary BHs fall within or above the mass gap (Column 10).

In Figure~\ref{fig:mass_comp} we show the primary and secondary component masses of all BBH mergers identified in our $f_{\rm acc} = 1$ cluster models, along with the GWTC-4 events that include at least one component within or above the pair-instability mass gap (defined here as any BH with mass of $40.5\,M_{\odot}$ or more). Accreted BH mergers (blue circles) have median primary masses around $40\,M_\odot$ and extend the high-mass tail up to $\sim200\,M_\odot$, while 2G mergers (black circles) are confined below $\sim80\,M_\odot$. A small number ($\sim 1\%$) of mergers in our models populate the same region of parameter space as GW231123 \citep{GW231123}, with total masses exceeding $200\,M_\odot$.  

In Figure~\ref{fig:mass_comp} we also show the mass ratios of secondary to primary BHs in merging binaries with dashed gray lines. Hierarchical mergers preferentially produce systems with unequal component masses ($q \lesssim 0.75$), while accretion-grown BHs (which are not subject to GW recoil kicks) tend to pair with other massive BHs ($q \sim 1$) early in their host's lifetime. About half of all binaries are subsequently ejected through dynamical encounters, and then merge outside the cluster at later times (see Figure~\ref{fig:mbbh_total}). Consequently, the accretion-driven channel can populate GW231123-like systems, where both BHs lie within or above the pair-instability mass gap (top-right shaded region in Figure~\ref{fig:mass_comp}) and have nearly equal masses.

\subsection{Spin-Mass Correlations}

While our previous work \citep{Kiroglu_2025a} investigated BH spin up from BH--star collisions in post-processing, the present study extends this by self-consistently incorporating BH spin evolution through accretion within the \texttt{CMC} framework, allowing us to follow their full dynamical history post-accretion. 
A key result of this work is the emergence of a mass-dependent BH spin distribution driven by accretion. BHs with masses $\lesssim 150\,M_\odot$ generally undergo one or two coherent accretion episodes, spinning them up to near-maximal values ($\gtrsim 0.7$). In contrast, more massive BHs ($\gtrsim 150\,M_\odot$), which often form through repeated, randomly oriented accretion episodes, tend to retain lower spins—similar to the spin-down behavior seen in BH growth via hierarchical mergers \citep[e.g.,][]{Hughes_2003}.

We show this trend in Figure~\ref{fig:coll_history}, where we illustrate three example pathways through which massive BHs are produced via stellar mergers and accretion, leading to diverse final spin outcomes. In the first case, a very massive star assembled through runaway stellar mergers is engulfed by an initially nonspinning stellar-mass BH; the large angular momentum from this single, massive accretion event spins the BH up to near maximal rotation. In the second case, a runaway stellar merger product collapses to a massive BH that subsequently accretes a star of lower mass, producing a modest spin ($\chi = 0.4$). In the third pathway, a mass‑gap BH formed via runaway stellar mergers experiences multiple BH–star collisions and accretion episodes with random angular‑momentum orientations, driving the spin down to very low values ($\chi\lesssim0.1$) despite the high final mass.

\begin{figure}
    \centering
\includegraphics[width=1\columnwidth]{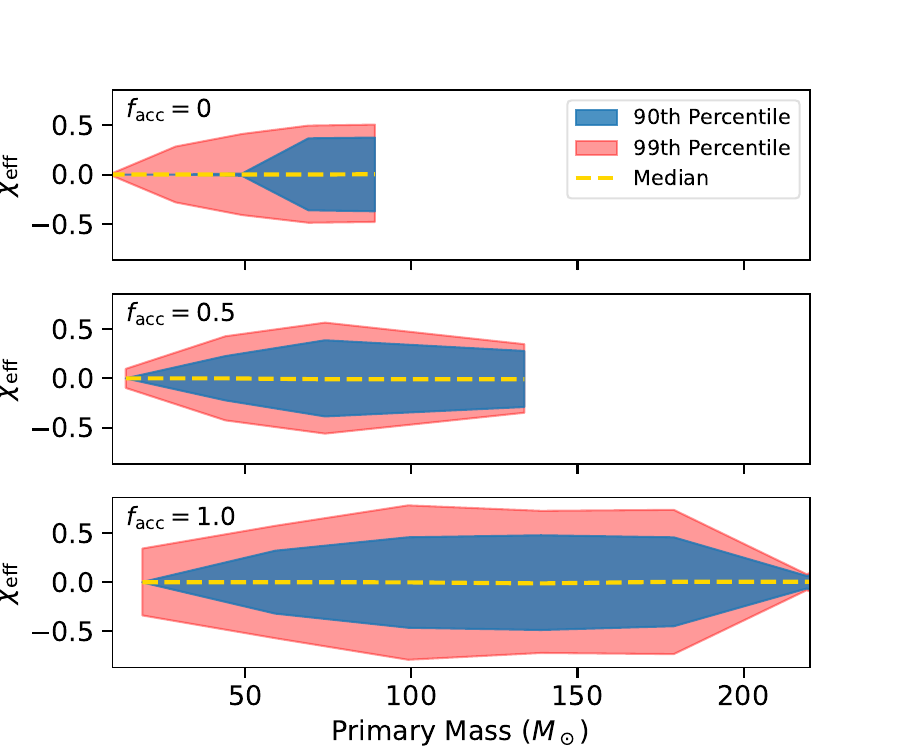}
    \caption{\footnotesize $\chi_{\mathrm{eff}}$ distributions for BBHs merging within a Hubble time are shown as a function of the primary BH mass for different accretion efficiencies, $f_{\rm acc}$. For each binary, $\chi_{\mathrm{eff}}$ is computed by averaging over $N=10^3$ realizations of randomly oriented spin vectors. Shaded regions indicate the 90th and 99th percentiles within each mass bin. The top panel shows the $\chi_{\mathrm{eff}}$ distribution when accretion is disabled, where broadening arises solely from hierarchical mergers. With accretion enabled, $\chi_{\mathrm{eff}}$ broadens with increasing mass up to $\sim100\,M_{\odot}$ and then narrows at higher masses due to spin-down from multiple randomly oriented accretion episodes, with the overall trend depending on $f_{\rm acc}$.
 }   
        \label{fig:chi_eff}
\end{figure}

In Figure~\ref{fig:chi_eff}, we show the effective spin distribution, $\chi_{\rm eff}$, as a function of primary BH mass for all merging BBHs in our models, grouped by accretion efficiency ($f_{\rm acc} = 0, 0.5, 1.0$). Although the distribution remains centered near zero due to isotropic spin orientations, its width grows with BH mass as we increase the accretion efficiency. As shown in our previous work \citep{Kiroglu_2025b}, some BBHs can develop preferential spin–orbit alignment following accretion; however, we neglect this effect here.

Increasing $f_{\rm acc}$ both populates the high-mass regime (up to $180\,M_\odot$) and enhances spin magnitudes, thereby broadening the overall $\chi_{\rm eff}$ distribution. This broadening eventually plateaus and begins to narrow at higher masses, as component BH spins decrease due to repeated, randomly oriented accretion episodes. For lower accretion efficiencies, this decline in $\chi_{\rm eff}$ occurs at smaller masses compared to the $f_{\rm acc}=1.0$ case, since producing more massive BHs requires a larger number of stochastic accretion events. This transition marks the regime where incoherent accretion dominates over coherent spin-up, leading to an overall spin-down of the most massive BHs. 
This behavior suggests that spin may serve as a powerful tracer of BH growth channels, with high spin ($\chi > 0.7$) massive BHs ($\approx 150\,M_{\odot}$) emerging as natural outcomes of accretion-driven growth in dense star clusters.

\section{Summary \& Discussion}
\label{sec:conclusion}

We have explored the formation pathways of massive, spinning BHs—particularly those within or above the pair-instability mass gap—and characterized their contribution to the population of observed BBH mergers.
We find that BHs can grow beyond $150\,M_\odot$ through successive stellar mergers and accretion, and their final spins are closely tied to their growth histories. 

Coherent accretion episodes spin up BHs efficiently, producing nearly maximally spinning BHs ($\chi \gtrsim 0.7$) for masses up to $\sim 150\,M_\odot$. Beyond this threshold, however, further mass growth typically requires multiple, randomly oriented accretion events, which tend to spin BHs down due to angular momentum cancellation. This leads to a distinct spin--mass correlation, where BH spins increase with mass up to the upper edge of the pair-instability mass gap but decline for the most massive BHs ($M \gtrsim 150\,M_\odot$), consistent with the high primary spin inferred for GW231123 ($\chi_1 \approx 0.9$ and $M_1 \approx 140\,M_\odot$).

The spin–mass trend from accretion is qualitatively similar to expectations from hierarchical mergers. BBH mergers involving 2G BHs yield higher-generation BHs that exhibit a broad range of spin values, depending sensitively on the spin alignment of their progenitors \citep[e.g.,][]{Rodriguez_2016,Rodriguez_2020,Atallah_2023,Fragione_2023}. However, merger products exceeding about $100\,M_\odot$ typically involve BHs of generation $>2$ and therefore generally exhibit spin values $\chi<0.7$ in massive clusters \citep[e.g.,][]{Mai2025}. In this context, the observed spins may pose a challenge for explaining GW231123-like systems with 3G or higher-generation mergers \citep[e.g.,][]{PassengerGW23hierarchical}. Beyond accretion from stars, alternative formation pathways that can produce high BH spins include accretion within active galactic nuclei (AGN) disks \citep[e.g.,][]{Tagawa2020,Vajpeyi2022,Bartos_2025}, chemically homogeneous evolution in close massive binaries \citep[e.g.,][]{Marchant2016,Marchant2024,Stegmann_2025}, and tidal spin-up in stellar binaries prior to collapse \citep[e.g.,][]{Bavera2021,Ma2023,Qin2023}.

Across all cluster models, about $20\%$ of BBH mergers involve at least one mass-gap BH ($M > 40.5\,M_\odot$), while only a few percent have both components in the mass gap. These BHs originate from a combination of stellar evolution of massive stellar collision products, accretion-driven growth, and hierarchical BBH mergers. Importantly, we show that such events can arise even in clusters with modest escape speeds, significantly broadening the range of environments capable of producing these extreme mergers.

While hierarchical mergers often produce binaries with unequal component masses, we find that accretion-grown massive BHs in clusters preferentially pair with other massive BHs at early times. The accretion-driven channel thus naturally explains GW231123-like systems, where both BHs lie within or above the pair-instability mass gap, have a mass ratio of $q \approx 0.75$, and exhibit high spins ($\chi \gtrsim 0.7$). Notably, recent population-inference work using LVK O4 data \citep{Ray2025} finds that the high-mass, broad-spin sub-population contains a significant fraction (roughly $50\%$) of systems with high mass ratios q=0.6–1, consistent with a substantial contribution from nearly equal-mass mergers, as expected from our accretion-driven channel.

Several recent studies propose different origins for GW231123, including coherent gas accretion in AGN disks or Population III remnants \citep[e.g.,][]{Bartos_2025}, the collapse massive stars \citep[e.g.,][]{Baumgarte2025,croon2025,Gottlieb2025,Torniamenti2025}, assembly in Population III star clusters \citep[e.g.,][]{Liu2025}, repeated mergers of stellar-mass BHs \citep[e.g.,][]{Li_2025,Mai2025,Stegmann_2025} and dynamical formation in AGN disks \citep[e.g.,][]{Delfavero2025}. Our findings indicate that spin–mass correlations provide a powerful observational diagnostic of the environments where massive BH binaries assemble, enabling us to distinguish between different mechanisms of BH growth. Moreover, evidence for a correlation between the effective inspiral spin and mass ratio can serve as an additional probe of binary formation pathways \citep[e.g.,][]{Callister_2021}. As the GW catalog continues to expand, constraints on these correlations will become increasingly robust, allowing us to disentangle the observational signatures of accretion-driven growth, hierarchical mergers, and binary evolution channels.

The accretion processes that shape GW sources may also produce bright electromagnetic transients during BH–star interactions, including ultra-long gamma-ray bursts \citep[e.g.,][]{Perets_2016,Beniamini2025}, X-ray flares \citep[e.g.,][]{Kremer_2022,Ryu_2022,Kiroglu_2023}, and wind–reprocessed optical/UV transients \citep[e.g.,][]{Kremer_2023}. Such events could be detectable in upcoming all-sky surveys such as the Vera C.\ Rubin Observatory and ULTRASAT, with expected rates of up to thousands per year from dense star clusters out to distances of several Gpc \citep[e.g.,][]{Kremer_2023,Rastello2025}.

This work represents a first self-consistent step toward understanding BH growth through collisions and accretion in dense star clusters using \texttt{\texttt{CMC}}. Our study is based on a limited number of cluster models, all placed at a fixed galactocentric distance and with a single metallicity. Among the various cluster properties, metallicity is expected to have the strongest impact on our results, as it influences stellar-wind mass loss, and thus the mass spectrum of stellar remnants, especially massive BHs. In contrast, variations in the cluster’s orbital location or present-day structure are less critical for the physical processes explored here.  A more comprehensive parameter study, spanning a broader range of metallicities and cluster initial conditions, will be presented in future work.

\begin{acknowledgements}

We thank Maya Fishbach, Claire Ye, and Jamie Lombardi for useful discussions. This work was supported by NSF grants AST-2108624 and AST-2511543, and NASA ATP grant 80NSSC24K0687. FK acknowledges support from a CIERA Postdoctoral Fellowship. We also acknowledge the computational resources and staff contributions provided for the \texttt{Quest} high-performance computing facility at Northwestern University. \texttt{Quest} is jointly supported by the Office of the Provost, the Office for Research, and Northwestern University Information Technology. F.A.R.\ acknowledges helpful discussions that took place at the Aspen Center for Physics, which is supported by NSF grant PHY-2210452. 
\end{acknowledgements}

\bibliographystyle{aasjournal}
\bibliography{mybib}

\end{document}